\documentclass[preprint,
prb]{revtex4}

\usepackage{amsmath}

\newcommand{{\bff}}{\boldsymbol{f}}

\newcommand{\gradv}{\boldsymbol{\nabla}}

\def\v#1{{\bf#1}}

\begin{document}

\title{How to obtain the covariant form of Maxwell's equations \\from the continuity equation}

\author{Jos\'e A. Heras}
\affiliation{Departamento de Ciencias B\'asicas, Universidad Aut\'onoma Metropolitana, Unidad Azcapotzalco, Av. San Pablo No. 180, Col. Reynosa, 02200, M\'exico D. F. M\'exico and Departamento de F\'isica y Matem\'aticas, Universidad Iberoamericana, Prolongaci\'on Paseo de la Reforma 880, M\'exico D. F. 01210, M\'exico}

\begin{abstract}
The covariant Maxwell equations are derived from the continuity equation for the electric charge. 
This result provides an axiomatic approach to Maxwell's equations in which charge conservation is emphasized as the fundamental axiom underlying these equations.  
\end{abstract}
\maketitle
\noindent{\bf 1. Introduction}
\vskip10pt
In classical electrodynamics the conservation of electric charge, which is expressed through the continuity equation 
\begin{align}
\gradv\cdot \v J+\frac{\partial\rho}{\partial t}=0, 
\end{align}
is not usually considered to be an independent assumption but a formal consequence of Maxwell's equations. The familiar argument is that the divergence of the Ampere-Maxwell law together with the Gauss law yield equation (1). Regarding equation (1) Griffiths has written [1]: ``This is, of course, the continuity equation  ---the precise mathematical statement of local conservation of charge. As I indicated earlier, it can be derived from Maxwell's equations. Conservation of charge is not an independent assumption, but a consequence of the laws of electrodynamics."  The argument is certainly more elegant in spacetime where the covariant form of the continuity equation reads $\partial_\nu J^\nu=0$. By applying the operator  $\partial_\nu$ to the inhomogeneous Maxwell equations (in Gaussian units) $\partial_\mu F^{\mu\nu} = (4\pi/c)J^\nu$ we obtain $\partial_\nu J^\nu=0$ because of the identity $\partial_\nu\partial_\mu F^{\mu\nu}\equiv 0$. 

Nevertheless, we have recently questioned [2] the traditional idea that charge conservation is a consequence of Maxwell's equations. In most textbooks 
the Ampere-Maxwell law, one of the Maxwell equations, is introduced by assuming the validity of the continuity equation. This fact involves a circular reasoning: to obtain the Ampere-Maxwell law we assume charge conservation and therefore this conservation law should not be considered as a consequence of Maxwell's equations. Furthermore, from a historical point of view charge conservation was the ingredient which allowed Maxwell to generalize the quasi-static Ampere's law: $\gradv\times \v B=(4\pi/c)\v J$ to the Ampere-Maxwell law: $\gradv \times \v B-(1/c)\partial\v E/\partial t=(4\pi/c)\v J.$ Accordingly, charge conservation should be interpreted as an axiom of Maxwell's equations rather than a consequence of them.

We have also demonstrated [2] an existence theorem for two retarded fields, which says that given localized time-dependent scalar and vector sources satisfying the continuity equation there exist two retarded vector fields that satisfy four field equations. 
When the sources are identified with the usual electromagnetic charge and current densities and appropriate choice of constants
made, the retarded vector fields become the generalized Coulomb and Biot-Savart laws [3,4] and their
associated field equations become Maxwell's equations. We have concluded that charge conservation is the fundamental axiom underlying Maxwell's equations. Our derivation of Maxwell's equations has been commented on by Jefimenko [5,6] and Kapu\`scik [7,8]. Colussi and Wickramasekara [9] 
have stressed the result that Maxwell's equations, which are Lorentz-invariant, have been obtained from the continuity equation, which is  Galilei-invariant according to these authors.

An advantage of the derivation of Maxwell's equations presented in reference 2 is that it naturally introduces the time-dependent extensions of the Coulomb and Biot-Savart laws in the form given by Jefimenko [3,4]. However, a practical disadvantage of such a derivation is that it involves unfamiliar and large identities involving derivatives of retarded quantities.

In this paper we extend our three-dimensional derivation of Maxwell's equations [2] to the four-dimensional spacetime. We
show how the covariant form of Maxwell's equations can be obtained from the continuity equation for the electric charge. The existence theorem for two retarded fields demonstrated in reference 2 is now formulated for a retarded antisymmetric field tensor in spacetime. The existence theorem in spacetime is then applied to the usual electromagnetic four-current and thus the electromagnetic field together with the covariant Maxwell equations are obtained. This four-dimensional derivation of Maxwell's equations is considerably shorter than that formulated in the three-dimensional space. Readers well versed in the mathematics of the spacetime of special relativity could find more simple and elegant the four-dimensional derivation because it does not involve large operations on explicit retarded quantities but instead it involves simplified tensor operations in which retardation is implicit in the retarded Green function of the free-space wave equation.

\vskip 15pt
\noindent{\bf 2. The covariant Maxwell equations}
\vskip 10pt
Greek indices $\mu, \nu, \kappa \ldots$ run from 0 to 3;  Latin indices $i,j,k,\ldots$ run from 1 to 3; $x=x^{\mu}=(ct,\v x)$ is the field point and $x'=x'^{\mu}=(ct',\v x')$ the source point; the signature of the metric of spacetime is $(+,-,-,-);$ $\varepsilon^{\mu\nu\alpha\beta}$ is the totally antisymmetric four-dimensional tensor with $\varepsilon^{0123}=1$ and $\varepsilon^{ijk}$ is the totally antisymmetric three-dimensional tensor with $\varepsilon^{123} = 1$. Summation convention on repeated indices is adopted.

The covariant form of Maxwell's equations can be written in Gaussian units as:
\begin{align}
\partial_\mu F^{\mu\nu} = \frac{4\pi}{c}J^\nu\quad{\rm and}\quad
\partial_\mu\!{^*}\!F^{\mu\nu} = 0,
\end{align}
where $F^{\mu\nu}$ is the electromagnetic field tensor and $^*\!F^{\mu\nu}=(1/2)\varepsilon^{\mu\nu\kappa\sigma}F_{\kappa\sigma}$ is the dual of 
$F^{\mu\nu}$. The antisymmetric tensor $F^{\mu\nu}$ is defined by its components $F^{i0}=(\v E)^i$ and $F^{ij}=-\varepsilon^{ijk}(\v B)_k,$
and therefore the components of $^*\!F^{\mu\nu}$ are $^*\!F^{i0}=(\v B)^i$ and $^*\!F^{ij}=\varepsilon^{ijk}(\v E)_k.$ 
The source of the tensor $F^{\mu\nu}$ is the electric four-current: $J^\nu =(c\rho,\v J),$
which is a conserved quantity
\begin{align}
\partial_\nu J^\nu =0.
\end{align}
As above mentioned, equation~(3) seems to be a formal consequence of  equations~(2) because of the identity $\partial_\nu\partial_\mu F^{\mu\nu}\equiv 0$, that is,  equations~(2) imply  equation~(3). We will investigate now the converse implication: To what extent  equation~(3) implies equations~(2)?
\vskip 15pt
\noindent{\bf 3. An existence theorem in spacetime}
\vskip 10pt
Let us formulate the following existence theorem: Given the localized four-vector  ${\cal J}^\nu$ satisfying the continuity equation
\begin{align}
\partial_\nu {\cal J}^\nu =0,
\end{align}
there exists the antisymmetric tensor field
\begin{equation}
{\cal F}^{\mu\nu}= \!\int\! d^4x' G(\partial'^\mu {\cal J}^\nu-\partial'^\nu {\cal J}^\mu),
\end{equation}
that satisfies the field equations
\begin{align}
\partial_{\mu} {\cal F}^{\mu\nu} = {\cal J}^\nu \quad{\rm and}\quad  
\partial_\mu\!{^*}\!{\cal F}^{\mu\nu} = 0,
\end{align}
where $G(x,x')=\delta(t'-t+R/c)/(4\pi R)$ with $R=|\v x-\v x'|$ is the retarded Green function of the free-space wave equation and $^*{\cal F}^{\mu\nu}=(1/2)\varepsilon^{\mu\nu\alpha\beta} {\cal F}_{\alpha\beta}$ is the dual of ${\cal F}^{\mu\nu}$. The function $G$ satisfies the free-space wave equation $\partial'_\mu\partial'^\mu G=\delta(x-x')$, where $\delta(x-x')$ is the four-dimensional delta function. The integral in  equation~(5) is over all spacetime. We note that $G$ is not an explicit Lorentz invariant object. This objection can be avoided by replacing $G$ by the less-known retarded invariant form $D(x,x')=\theta(x_0-x'_0)\delta[(x-x')^2]/(2\pi)$,
where $\theta$ is the theta function [10].

The proof of the theorem follows from the tensor identity
\begin{equation}
\partial_{\mu} G(\partial'^{\mu} {\cal J}^{\nu}-\partial'^{\nu} {\cal J}^{\mu})\!-\!\partial'_{\mu} (G\partial'^{\nu}{\cal J}^{\mu}-{\cal J}^{\nu}\partial'^{\mu} G)\!=\! {\cal J}^{\nu}\delta(x-x').
\end{equation}
This identity is of general validity for a four-vector ${\cal J}^\nu$ that satisfies equation (4) but otherwise is arbitrary. In particular,  equation~(7) is valid when ${\cal J}^\nu$ is localized into a finite region of spacetime, property that guaranties that surface integrals in spacetime involving ${\cal J}^\nu$ vanish at infinity. Equation (7) can be obtained from Eq.~(4) as follows. Assume that ${\cal J}^{\nu}$ satisfies the continuity equation evaluated at the source point: 
\begin{equation}
\partial'_{\mu} {\cal J}^\mu =0. 
\end{equation}
We apply the operator $-G\partial'^{\nu}$ to equation~(8) and add the term ${\cal J}^{\nu}\partial'_{\mu}\partial'^{\mu} G$ to both sides: 
\begin{equation}
-G\partial'{^\nu}\partial'_\mu {\cal J}^{\mu}+{\cal J}^{\nu}\partial'_{\mu}\partial'^{\mu} G={\cal J}^{\nu}\partial'_{\mu}\partial'^{\mu} G. 
\end{equation}
The left-hand side of equation (9) can be written as
\begin{equation}
-\partial'_{\mu} G(\partial'^{\mu} {\cal J}^{\nu} -\partial'^{\nu} {\cal J}^{\mu})-\partial'_\mu (G\partial'^\nu { \cal J}^\mu-{ J}^\nu\partial'^\mu G) ={\cal J}^{\nu}\partial'_{\mu}\partial'^{\mu} G.
\end{equation}
When $-\partial'_{\mu} G=\partial_{\mu} G$ is used in the first term of the left-hand side of equation~(10) and $\partial'_\mu\partial'^\mu G=\delta(x-x')$ in its right-hand side, we finally obtain equation~(7).

We integrate  equation~(7) over all spacetime
\begin{align}
\partial_\mu\! \int\! d^4x' G(\partial'^\mu {\cal J}^\nu -\partial'^\nu {\cal J}^\mu)- \int\! d^4 x'\partial'_\mu (G\partial'^\nu {\cal J}^\mu-{\cal J} ^\nu\partial'^\mu G)
=\! \int\!  d^4 x'{\cal J}^\nu\delta(x-x'),
\end{align}
where $\partial_\mu$ has been extracted from the first integral of the left-hand side. The second integral on this side can be transformed into a surface integral which vanishes at infinity because of ${\cal J} ^\nu$ is assumed to be a localized quantity. The term on the right-hand side can be integrated yielding ${\cal J} ^\nu$. Thus  equation (11) reduces to
\begin{equation}
\partial_\mu \!\int\! d^4x' G(\partial'^\mu {\cal J}^\nu-\partial'^\nu {\cal J}^\mu)={\cal J}^\nu.
\end{equation}
This equation  shows the existence of the antisymmetric tensor
\begin{equation}
{\cal F}^{\mu\nu}=\!\int \!d^4x' G(\partial'^\mu {\cal J} ^\nu-\partial'^\nu {\cal J} ^\mu),
\end{equation}
in terms of which  equation~(12) takes the compact (inhomogeneous) form:
\begin{equation}
\partial_\mu {\cal F}^{\mu\nu}={\cal J}^\nu.
\end{equation}
The dual of ${\cal F}^{\mu\nu}$ can be written as
\begin{equation}
^*{\cal F}^{\mu\nu}=\varepsilon^{\mu\nu\kappa\lambda}\! \int\! d^4x' G\partial'_{\kappa} {\cal J}_{\lambda}.
\end{equation}
We apply the operator $\partial_{\mu}$ to equation~(15) and perform an integration by parts to obtain
\begin{equation}
\partial_\mu{\!^*}\!{\cal F}^{\mu\nu}=\varepsilon^{\mu\nu\kappa\lambda}\!\int \!d^4x' G\partial'_\mu\partial'_{\kappa} {\cal J}_{\lambda}-\!\int\! d^4x'\partial'_\mu(\varepsilon^{\mu\nu\kappa\lambda}G\partial'_{\kappa} {\cal J}_{\lambda}).
\end{equation}
The first integral vanishes because of the symmetry of the indices $\mu$ and $\kappa$ in $\partial'_\mu\partial'_{\kappa}$ and the antisymmetry of such indices in $\varepsilon^{\mu\nu\kappa\lambda}$. The second integral can be transformed into a surface integral which vanishes at infinity. Then equation~(16) reduces to the compact (homogeneous) form
\begin{equation}
\partial_\mu{\!^*}\!{\cal F}^{\mu\nu}=0.
\end{equation}
Once equations~(13), (14), and (17) have been obtained, the theorem has been demonstrated. Notice that this theorem applies to any conserved four-vector which does not necessarily belong to the electromagnetic theory. 

\vskip 15pt
\noindent{\bf 4. Deriving Maxwell's equations}
\vskip 10pt
Let us apply the existence theorem to the four-vector ${\cal J}^\mu=\beta\!J^\mu$, where $J^\mu=(c\rho,\v J)$ is the usual electric four-current with $\rho$ and $\v J$ being the charge and current densities and $\beta$ is a constant. Then there exists the tensor field 
\begin{equation}
{\cal G}^{\mu\nu}=\beta\!\int \!d^4x' G(\partial'^\mu J^\nu-\partial'^\nu J^\mu),
\end{equation}
which satisfies the field equations
\begin{align}
\partial_\mu {\cal G}^{\mu\nu} = \beta J^\nu  \quad{\rm and}\quad   
 \partial_\mu{\!^*}\!{\cal G}^{\mu\nu} = 0,
\end{align}
where $^*{\cal G}^{\mu\nu}$ can be written as
\begin{equation}
^*{\cal G}^{\mu\nu}=\beta \varepsilon^{\mu\nu\kappa\lambda} \!\int \!d^4x' G\partial'_{\kappa} J_{\lambda}.
\end{equation}
The constant $\beta$ belongs to the $\alpha\beta\gamma$-system which allows us to write electromagnetic expressions in a way that is independent of units [11]. The constants $\alpha,\beta$ and $\gamma$ satisfy the relation $\alpha=\beta\gamma c^{\:2}$ with $c$ being the speed of light in vacuum [11].
 If we let $\alpha=1/\epsilon_0, \beta=\mu_0$, and $\gamma=1$, we obtain expressions in SI units, and if we let $\alpha=4\pi, \beta=4\pi/c$, and $\gamma=1/c$, we obtain expressions in Gaussian units.

To verify that equations~(19) are already the covariant Maxwell equations, that is, that the tensor ${\cal G}^{\mu\nu}$ identifies with the well-known electromagnetic field tensor $F^{\mu\nu},$ we proceed as follows. Let us label the components of the tensor ${\cal G}^{\mu\nu}$ in equation (18) as ${\cal G}^{i0}=(\beta c/\alpha)(\v X)^i$ and ${\cal G}^{ij}=-\varepsilon^{ijk}(\v Y)_k$, where $\v X$ and $\v Y$ are two time-dependent vector fields. The components of the dual tensor $^*{\cal G}^{\mu\nu}$ can be obtained from those of ${\cal G}^{\mu\nu}$ by making the dual changes $(\beta c/\alpha)(\v X)^i\rightarrow (\v Y)^i$ and $(\v Y)_k \rightarrow -[1/(\gamma c)](\v X)_k$. Notice that $\beta c/\alpha=1/(\gamma c)$. It follows that $^*{\cal G}^{i0}=(\v Y)^i$ and $^*{\cal G}^{ij}=[1/(\gamma c)]\varepsilon^{ijk}(\v X)_k$. Our plan is now to identify the vectors $\v X$ and $\v Y$ with the electric and magnetic fields $\v E$ and $\v B$. With this purpose in mind, let us consider the $i0$ component of equation (18) which can be written as
\begin{align}
(\v X)^i&=\frac{\alpha}{4\pi}\int \!d^4x' G(\partial'^i J^0-\partial'^0 J^i),\nonumber\\
&= \frac{\alpha}{4\pi}\int \!\int\! d^3x'dt' G\bigg\{\!-\!(\gradv'\rho)^i\!-\!\frac{1}{c^2}\bigg(\frac{\partial\v J}{\partial t'}\bigg)^i\bigg\}\nonumber\\
&= \frac{\alpha}{4\pi}\int\! d^3x'\frac{1}{R}\bigg[\!-\!\gradv'\rho\!-\!\frac{1}{c^2}\frac{\partial\v J}{\partial t'}\;\bigg]^i_{\rm ret},
\end{align}
where the square brackets $[\;]_{\rm ret}$ indicate that the enclosed quantity is to be evaluated at the retarded time $t'=t-R/c.$ The right-hand side of Eq.~(21) is the same as the right-hand side of the i-component of the retarded electric field 
\begin{align}
(\v E)^i= \frac{\alpha}{4\pi}\int \!d^3x'\frac{1}{R}\bigg[\!-\!\gradv'\rho\!-\!\frac{1}{c^2}\frac{\partial\v J}{\partial t'}\;\bigg]^i_{\rm ret}.
\end{align}
Comparison between equations (21) and (22) yields the identification $\v X=\v E$. We consider now the $i0$ component of equation (20) which can be expressed as 
\begin{align}
(\v Y)^i=& \frac{\beta}{4\pi}\epsilon^{i0\alpha\beta}\int \!d^4x' G\partial'_{\alpha} J_{\beta},\nonumber\\
=& \frac{\beta}{4\pi}\int\!\int \!d^3x'dt' G\{\epsilon^{0ijk}\partial'_j J_k\}\nonumber\\
=& \frac{\beta}{4\pi}\int \!d^3x'\frac{1}{R}[\gradv'\times \v J]^i_{\rm ret}.
\end{align}
The right-hand side of equation~(23) is the same that the right-hand side of the i-component of the retarded magnetic field 
\begin{align}
(\v B)^i=  \frac{\beta}{4\pi}\int\! d^3x'\frac{1}{R}[\gradv'\times \v J]^i_{\rm ret}.
\end{align}
Comparison between equations (22) and (23) yields the identification $\v Y=\v B$. Then the tensor ${\cal G}^{\mu\nu}$ naturally identifies with the electromagnetic field tensor  
\begin{equation}
F^{\mu\nu}=\beta\!\int \!d^4x' G(\partial'^\mu J^\nu-\partial'^\nu J^\mu),
\end{equation}
which satisfies the Maxwell equations
\begin{align}
\partial_\mu F^{\mu\nu} = \beta J^\nu  \quad{\rm and}\quad   
 \partial_\mu{\!^*}\!F^{\mu\nu} = 0,
\end{align}
where $^*\!F^{\mu\nu}$ can be written as
\begin{equation}
^*\!F^{\mu\nu}=\beta \varepsilon^{\mu\nu\kappa\lambda} \!\int \!d^4x' G\partial'_{\kappa} J_{\lambda}.
\end{equation}
The components of $F^{\mu\nu}$ are given by 
\begin{equation}
F^{i0}=\frac{\beta c}{\alpha}(\v E)^i \quad{\rm and}\quad F^{ij}=-\varepsilon^{ijk}(\v B)_k,
\end{equation}
where $(\v E)^i$ and $(\v B)_k$ are the components of the electric and magnetic fields. It follows that 
\begin{align}
^*\!F^{i0}=(\v B)^i\quad {\rm and }\quad
^*\!F^{ij}=\frac{1}{\gamma c }\varepsilon^{ijk}(\v E)_k.
\end{align}
Therefore, the covariant form of Maxwell's equations have been obtained as an application of the existence theorem formulated here.

For completeness, we can verify that Eqs.~(26) yield the three-dimensional form of Maxwell's equations in $\alpha\beta\gamma$-units. 
By using equations~(28), (29) and $\partial_{\mu}=[(1/c)\partial/\partial t,\gradv]$, we can write the four-vectors [11]:
\begin{align}
\partial_{\mu}F^{\mu\nu}&=\bigg (\!\frac{\beta c}{\alpha}\nabla \cdot \v E,\;\nabla \times \v B-\frac{\beta}{\alpha}\frac{\partial \v E}{\partial t} \bigg ),\\
\partial_{\mu}\!{^*}\!F^{\mu\nu}&=\bigg (\!\nabla \cdot \v B, \; -\frac{1}{\gamma c}\nabla \times \v E-\frac 1c\frac{\partial \v B}{\partial t} \bigg).
\end{align}
From  equations~(26), (30), (31) and $J^\nu=(c\rho, \v J),$ we obtain the three-dimensional form of Maxwell's equations in $\alpha\beta\gamma$ units 
for sources in vacuum [11]:
\begin{align}
\gradv\cdot\v E&=\alpha\rho,\\
\gradv\cdot\v B&= 0,\\
\gradv\times \v E+\gamma\frac{\partial \v B}{\partial t}&= 0,\\
\gradv\times \v B-\frac{\beta}{\alpha}\frac{\partial \v E}{\partial t} &=\beta\v J.
\end{align}
In particular, if we let $\alpha=1/\epsilon_0, \beta=\mu_0$, and $\gamma=1$, then we obtain Maxwell's equations in SI units,

\vskip 15pt
\noindent{\bf 5. Discussion}
\vskip 10pt
In section 4 we have established the identification ${\cal G}^{\mu\nu}=F^{\mu\nu}$ on the basis that equations (22) and (24) are already known. But suppose that we do not know these equations but only the experimental time-independent Coulomb and Biot-Savart laws. In this case the time-independent limit of the $i0$ component of  equation~(18) can be written as
\begin{equation}
(\v X)^i= \frac{\alpha}{4\pi}\!\int d^3x'\frac{(\hat{\v R})^i\rho}{R^2},
\end{equation}
where $\hat{\v R}={\v R}/R$. From the experimental law $\bff =q_0\v E$ for the force $\bff$ acting on a charge $q_0$ we know that there exists the electrostatic field $\v E$ with components
\begin{equation}
(\v E)^i=\frac{\alpha}{4\pi}\!\int\! d^3x'\frac{(\hat{\v R})^i\rho}{R^2}.
\end{equation}
Comparison between equations~(36) and (37) yields the identification $\v X=\v E$ in the time-independent regime.  We consider now the
time-independent limit of the $i0$ component of equation~(20) which can be expressed as
\begin{equation}
(\v Y)^i = \frac{\beta}{4\pi}\!\int d^3x'\! \bigg(\v J\times\!\frac{\hat{\v R}}{R^2 }\bigg)^i.
\end{equation}
From the experimental law $\bff=\gamma\!\int d^3x'\v J_0\times \v B$ for the force $\bff$ acting on a current $\v J_0$ we know that there exists the magnetostatic field $\v B$ given by
\begin{equation}
(\v B)^i= \frac{\beta}{4\pi}\!\int d^3x'\! \bigg(\v J\times\!\frac{\hat{\v R}}{R^2 }\bigg)^i.
\end{equation}
Comparison between equations~(38) and (39) leads to the identification $\v Y=\v B$ in the time-independent regime. 

Therefore, by requiring that the time-independent limit of  equations~(18) and (20) be consistent with the experimental Coulomb and Biot-Savart laws, we obtain the identifications $\v X=\v E$ and $\v Y=\v B$ in the time-independent regime. Such identifications naturally extend to the time-dependent regime of the theory and so we conclude again the identification ${\cal G}^{\mu\nu}=F^{\mu\nu}.$
Let us imagine for a moment that there is a physicist that only knows the experimental Coulomb and Biot-Savart laws. He then can infer the electromagnetic field together with Maxwell's equations by assuming the validity of the continuity equation. The physicist would correctly conclude that charge conservation is the fundamental postulate underlying Maxwell's equations.

It can be argued that the derivation of the covariant form of Maxwell's equations from the continuity equation involves other implicit physical postulates like retardation (causality) which is implicit in the use of the retarded Green function $G$. If instead of the Minkowski spacetime 
with metric signature $(+,-,-,-)$ together with its associated retarded Green function $G=\delta\{t'-t+R/c\}/(4\pi R)$, we consider, 
for example, an Euclidean four-space [12-14] with metric signature $(+,+,+,+)$ together with its associated retarded-imaginary Green function [14] $G_I=\delta\{t'-t+R/(ic)\}/(4\pi R)$, then we obtain a set of Maxwell-like field equations in the 
Euclidean four-space [12-14]. This means that the continuity equation can also imply other electromagnetic theories different from that of Maxwell. 
This is an expected result because charge conservation is independent of the signature of the metric of spacetime. The independence of the signature can be verified in the equation 
\begin{equation}
\partial_\alpha J^\alpha=\partial^\alpha J_\alpha=\gradv\cdot\v J+\frac{\partial \rho}{\partial t}=\hat{\partial}_\alpha \hat{J}^\alpha=\hat{\partial}^\alpha \hat{J}_\alpha,
\end{equation}
where $\partial_{\alpha}=[(1/c)\partial/\partial t,\gradv],\;\partial^{\alpha}=[(1/c)\partial/\partial t,-\gradv],\;J_\alpha=(c\rho,-\v J)$ and $J^\alpha=(c\rho,\v J)$ are objects in the Minkowski spacetime and $\hat{\partial}_{\alpha}=[(1/c)\partial/\partial t,\gradv],\;\hat{\partial}^{\alpha}=[(1/c)\partial/\partial t,\gradv],\; \hat{J}_\alpha=(c\rho,\v J)$
 and $\hat{J}^\alpha=(c\rho,\v J)$ are objects in the Euclidean spacetime (there is no difference between covariance and contravariance in this spacetime). Hehl and Obukhov [15] have pointed out that charge conservation is metric-independent because it is based
on a counting procedure for elementary charges. Therefore if one first postulates the validity of the continuity equation in any four-space, the particular use of the Minkowski spacetime (together with its associated retarded Green function $G$) would not really be a new postulate, but only just a particular application of the initial postulate.                                                    

\vskip 15pt
\noindent{\bf 6. The subtle relation between sources and fields: Is this an egg-hen problem?}
\vskip 10pt
A referee has pointed out that $\partial\rho/\partial t$ and $\v J$ can only be produced by electric and magnetic fields (take Ohm's law for instance $\v J = \sigma \v E).$ Therefore a circular process seems to be unavoidable in electromagnetism: $\rho$ and $\v J$  imply $\v E$ and $\v B$ which in turn imply new $\rho_1$ and $\v J_1$, and so on. Because of this circular characteristic, it is not clear if $\v E$ and $\v B$ (satisfying Maxwell's equations) are a consequence of $\rho$ and $\v J$ (satisfying the continuity equation) or vice versa.  According to the referee it seems a matter of taste to say which one is a consequence of the other. In other words: From referee's comment we could conclude that the connection between sources and fields is a little bit like the egg-hen problem: Who was first? 

Let us to briefly discuss the very deep relation between sources  and fields in the context of the electromagnetic theory.
 It is now a common place to say that charges and currents are the causes of the electric and magnetic fields, but this was not so in the 19th century [16,17]. Two rival point of views on what is now known as electromagnetic theory coexisted through most of the 19th century. On one hand, the field point of view followed by Faraday, Maxwell, Thomson, Fitzgerald, and Larmor who considered the fields (and their ``observable" lines of force) to be the primary objects of the theory. They viewed the charges as manifestations of the terminal points of lines of force having no independent or substantial existence [17]. On the other hand, the charge-interaction point of view followed by Ampere, Neumann and Weber who considered  
the electromagnetic phenomena as originating in the interaction of charges which was regarded by some as mediated by an ethereal continuum and by others as a direct action at a distance. In the charge-interaction view, charges and currents were the primary objects of the theory and the causes of the forces. 
Modern field theory of the 20th century resulted from a combination of these two point of views. Charges were regarded as sources of the electromagnetic forces (this was the legacy of the charge-interaction point of view) which were considered as propagating through the Maxwellian field (this was the legacy of the field point of view). 

The existence of a causal relation between sources and fields is now universally accepted in electromagnetism: $\rho$ and $\v J$ (causes) produce the fields $\v E$ and $\v B$ (effects). This relation is clearly expressed by the retarded solutions of Maxwell's equations in the form of time-dependent extensions of the Coulomb and Biot-Savart laws (in Gaussian units) [3]:	
\begin{align}
\v E(\v x,t)&\!=\!\int \!d^3x'\;\frac{\rho(\v x',t\!-\!R/c)\hat{\v R}}{R^2}+\frac{\partial}{\partial t}\!\int\! d^3x'\bigg(\!\frac{\rho(\v x',t\!-\!R/c)\hat{\v R}}{Rc}-\frac{\v J(\v x',t\!-\!R/c)}{Rc^2}\!\bigg),\\
\v B(\v x,t)&\!=\!\int\! d^3x'\; \frac{\v J(\v x',t\!-\!R/c)\! \times\!\hat{\v R}}{R^2c }
+\frac{\partial}{\partial t}\!\int\! d^3x'\;\frac{\v J(\v x',t\!-\!R/c) \times\hat{\v R}}{R c^2}.
\end{align}
According to these equations the values of $\v E$ and $\v B$ at the instant $t$ are determined by the values of $\rho$ and $\v J$ at the retarded (previous) time $t\!-\!R/c$. This result  clearly agrees with the principle of causality which roughly speaking says that the causes (in this case $\rho$ and $\v J$) precede in time to their effects (in this case $\v E$ and $\v B$). Traditionally we place the causes on the right-hand side of equations and the effects on their left-hand side. 

The principle of causality provides then a reasonable argument to elucidate what are the causes and what are the effects in electromagnetism.  Who were first? By looking at Eqs.~(41) and (42), our answer is straightforward: $\rho$ and $\v J$ occurred first and therefore they should be considered the causes of $\v E$ and $\v B$. This interpretation would seem incontrovertible ---and so the supposed egg-hen problem involved in the source-field relation would be an illusion--- except because there exist also the so-called advanced solutions of Maxwell's equations, which can be written as
\begin{align}
\int \!d^3x'\frac{\rho(\v x',t\!+\!R/c)\hat{\v R}}{R^2}\!-\!\frac{\partial}{\partial t}\!\int\! d^3x'\!\bigg(\!\frac{\rho(\v x',t\!+\!R/c)\hat{\v R}}{Rc}\!+\!\frac{\v J(\v x',t\!+\!R/c)}{Rc^2}\!\bigg)&=\v E(\v x,t),\\
\int\! d^3x' \frac{\v J(\v x',t\!+\!R/c)\! \times\!\hat{\v R}}{R^2c }
\!-\!\frac{\partial}{\partial t}\!\int\! d^3x'\frac{\v J(\v x',t\!+\!R/c) \times\hat{\v R}}{R c^2}&=\v B(\v x,t).
\end{align}
According to these equations the values of $\v E$ and $\v B$ at the instant $t$ are related with the values of $\rho$ and $\v J$ at the advanced (posterior) time $t\!+\!R/c$. We again invoke the principle of causality to determine what are the causes and the effects in electromagnetism.  Who were first? By looking at Eqs.~(43) and (44) we note that $\v E$ and $\v B$ occurred first and therefore they should be considered the causes of $\rho$ and $\v J$. Furthermore, if we introduce the variable $\tau=t+R/c$ then Eqs.~(43) and (44) can alternatively be written as
\begin{align}
\int \!d^3x'\frac{\rho(\v x',\tau )\hat{\v R}}{R^2}\!-\!\frac{\partial}{\partial t}\!\int\! d^3x'\!\bigg(\!\frac{\rho(\v x',\tau)\hat{\v R}}{Rc}\!+\!\frac{\v J(\v x',\tau)}{Rc^2}\!\bigg)&=\v E(\v x,\tau\!-\!R/c),\\
\int\! d^3x' \frac{\v J(\v x',\tau)\! \times\!\hat{\v R}}{R^2c}
\!-\!\frac{\partial}{\partial t}\!\int\! d^3x'\frac{\v J(\v x',\tau) \times\hat{\v R}}{R c^2}\!&=\!\v B(\v x,\tau\!-\!R/c),
\end{align}
These equations say that the values of $\rho$ and $\v J$ at the instant $\tau$ are related with the values of $\v E$ and $\v B$ at the previous time  
$\tau\!-\!R/c$. In the light of the causality principle, the advanced solutions of Maxwell's equations are consistent with the idea that $\v E$ and $\v B$ are the causes of $\rho$ and $\v J,$ but the retarded solutions of these same equations are consistent with the idea that $\rho$ and $\v J$ are the causes of $\v E$ and $\v B$. As above mentioned, the subtle relation between sources and fields in electromagnetism is a little bit like the egg-hen problem.

It can be argued, however, that the above analysis is purely formal and that other physical considerations allow us to identify the causes and the effects in electromagnetism. Suppose we have an electron at rest which is associated with an electrostatic field. In this case, we have neither retarded nor advanced times and so we cannot clearly distinguish between causes and effects. Suppose now that the electron is accelerated by mechanical forces. The resulting electric current obeys the continuity equation and the reason the charge moves has nothing to do with electromagnetism. Experience shows that the electron radiates, according to Maxwell's equations, and that this radiation carries of electromagnetic energy, which must come at the expense of the particle's mechanical energy. The radiated energy detaches from the electron, propagates off to infinity never coming back and we can physically detect this electromagnetic energy. In this dynamical case the radiation fields clearly arise as a consequence of the motion of charges which was originated by mechanical and not electromagnetic means. This simple example shows that the continuity equation is valid outside the domain of Maxwell's equations and therefore it should be regarded as a stronger axiom.

\vskip 15pt
\noindent{\bf 7. Concluding remarks}
\vskip 10pt
R. G. Brown wrote [18]: ``... observe that if we take the 4-divergence of both sides of the inhomogeneous
Maxwell equations:   $\partial^\alpha\partial^\beta F_{\beta\alpha}=\mu_0\partial^\alpha J_\alpha=0$ the left hand side vanishes because
again, a symmetric differential operator is contracted with a completely antisymmetric field
strength tensor. Thus $\partial^\alpha J_\alpha=0$, which, by some strange coincidence, is the charge-current
conservation equation in four dimensions. Do you get the feeling that something very deep
is going on? This is what I love about physics. Beautiful things are really beautiful!" I think there is no strange coincidence. This is what I love about physics! The beautiful equation $\partial^\alpha J_\alpha=0$ representing charge conservation is the fundamental symmetry behind the inhomogeneous and homogeneous Maxwell equations. ``That something very deep" is just charge conservation. To consider the continuity equation as the basic axiom to build the set of equations known as Maxwell's equations is then a very natural idea. In this paper we have followed this idea by obtaining the covariant form of Maxwell's equations from the covariant form of the continuity equation. 

It can be argued that other physical assumptions like the use of the Minkowski spacetime with its speed of light $c$
and its associated retarded Green function $G$ are ingredients implicit in the covariant derivation of Maxwell' equations presented here. However, if we first postulate the validity of the continuity equation in any four-space, the particular use of the Minkowski spacetime (together with its associated retarded Green function $G$ and speed $c$) would not really be a new postulate, but merely an application of our initial postulate.

We advocate an axiomatic presentation of Maxwell's equations in undergraduate and graduate courses of electromagnetism. According to Obradovic [19]: ``The axiomatic method
offers the shortest way to the essence of any theory, enables its more accurate formulation
and more profound and complete interpretation." In the case of Maxwell's equations the essence is charge conservation which should be considered the fundamental postulate for an axiomatic presentation of these equations. In the practice, undergraduate instructors can find the basis for this presentation in equations (4a) and (4b) of Ref. 2 and graduate instructors in equation (7) of this paper. \\

\noindent {\bf Acknowledgment}

Useful comments by an anonymous referee are gratefully acknowledged.

{}
\end{document}